\documentclass[%
letterpaper,aps,prl,superscriptaddress,showpacs,reprint,amsmath
]{revtex4-1}

\usepackage{graphicx}

\newcommand{\ket}[1]{| #1 \rangle}
\newcommand{\bra}[1]{\langle #1 |}
\newcommand{\braket}[1]{\langle #1 \rangle}
\newcommand{\moment}[1]{\langle #1 \rangle}
\newcommand{\Tr}{\operatorname{Tr}}

\newcommand{\UT}{%
Department of Applied Physics, School of Engineering, \\
The University of Tokyo, 7-3-1 Hongo, Bunkyo-ku, Tokyo 113-8656, Japan
}
\newcommand{\UP}{%
Department of Optics, Palack\'y University,
17. listopadu 1192/12, 77146 Olomouc, Czech Republic
}

\begin{document}
\title{%
Generation and characterization of resource state
for nonlinear cubic phase gate
}

\author{Mitsuyoshi Yukawa}
\affiliation{\UT}
\author{Kazunori Miyata}
\affiliation{\UT}
\author{Hidehiro Yonezawa}
\affiliation{\UT}
\author{Petr Marek}
\affiliation{\UP}
\author{Radim Filip}
\affiliation{\UP}
\author{Akira Furusawa}
\affiliation{\UT}

\begin{abstract}
Unitary non-Gaussian nonlinearity is one of the key components
required for quantum computation and other developing applications of
quantum information processing.
Sufficient operation of this kind is still not available,
but it can be approximatively implemented with
help of a specifically engineered resource state constructed from
individual photons.
We present experimental realization and thorough analysis of
such quantum resource state,
and confirm that the state does indeed possess
properties of a state produced by unitary dynamics
driven by cubic nonlinearity.
\end{abstract}

%03.67.Lx Quantum computation architectures and implementations
%42.50.Dv Quantum state engineering and measurements
%42.50.Ex Optical implementations of quantum information
%         processing and transfer
%42.65.-k Nonlinear optics

\maketitle

%\emph{Introduction} -
Nonlinear interactions capable of
manipulating quantum state of the harmonic oscillators
form a very challenging area of the recent development
in the field of modern quantum physics.
Handling these interactions is necessary
not only for the understanding of
quantum nonlinear dynamics of the harmonic oscillators, but also
for achieving the standing long-term goal of quantum information -
the universal quantum computation \cite{comp,compcluster}.
The operations needed are unitary,
and both Gaussian and non-Gaussian \cite{Akirabook}.
For a harmonic oscillator
representing a single mode of electro-magnetic radiation,
Gaussian operations are relatively easy to obtain,
but unitary non-Gaussian nonlinearities
performed on the oscillator alone are either not available,
or too weak to have an observable quantum effect.
For other physical systems,
such as cold atoms \cite{coldatoms} or trapped ions \cite{trappedions},
the non-Gaussian operations could be implemented
using additional an-harmonic potentials,
but Gaussian operations between the oscillators are hard to come by.

Probabilistic quantum nonlinear operations for light can be, in principle,
obtained by letting the system interact with
individual atoms, ions \cite{atoms} or
similar solid-state physical systems \cite{solid},
and measuring the discrete system afterwards.
This inherently probabilistic way of implementing nonlinearity
is more suitable for state preparation than for state processing,
but this issue can be circumvented
by employing the paradigm of
ancilla-and-measurement induced operations \cite{cubic,Marek11}.
However, this requires access to deterministic Gaussian processing
in the form of operations and measurements,
which is limited for cavity fields used in interactions with these systems.
That is quite unfortunate,
because extremely fine experimental control of those systems
was demonstrated by generating superpositions of
large amplitude coherent states \cite{microcats}.
These quantum states possess strong nonlinear properties,
and they were not previously observed in the trapped ions \cite{ioncats},
the circuit cavity electrodynamics \cite{circuitcats},
and in the optical experiments with traveling light \cite{optcat1,optcat2}.
The last mentioned optical system has one distinct advantage, though.
It allows easy implementation of any Gaussian operation and
Gaussian measurement \cite{mio1,mio2,mio3,mio4},
and non-Gaussian nonlinearity can be obtained by
employing single photon detectors \cite{Akirabook}.
Recently, this approach has been used
to generate quantum states with nonlinear properties
comparable to those of atomic and solid-state systems \cite{optcat3}.
The toolbox of traveling light quantum optics therefore contains
probabilistic highly non-linear operations
as well as deterministic Gaussian operations,
which makes this platform very promising for
tests of unitary nonlinear dynamics.

In principle, to realize an arbitrary unitary operation of
quantum harmonic oscillator,
it is sufficient to have access to the quantum cubic nonlinearity
\cite{comp,Loock11}.
Cubic nonlinearity is represented by Hamiltonian
$\hat{H} \propto \hat{x}^3$ \cite{cubic},
where $\hat{x} = (\hat{a} + \hat{a}^{\dag})/\sqrt{2}$
is the position operator of the quantum harmonic oscillator
($\hat{a}$ is the annihilation operator).
The evolution driven by this Hamiltonian preserves the behavior of $\hat{x}$,
while changing the complementary momentum quadrature $\hat{p}$
by an additive term proportional to $\hat{x}^2$.
As of now, neither quantum cubic nonlinearity,
nor quantum states produced by it (cubic states),
have been observed on any experimental platform.
Beginning from a ground state,
even the weak cubic interaction generates
highly nonclassical states \cite{Ghose07}.
However, nonclassicality of these states lies in the superposition of
$\ket{1}$ and $\ket{3}$ ($\ket{1 \& 3}$ for shorthand)
and it is unfortunately masked by
the superposition of $\ket{1 \& 3}$
with the dominant ground state $\ket{0}$ \cite{Marek11},
especially considering its fragility
with regards to damping of the oscillator.
It is therefore challenging
not only to generate and detect these states
but also to understand and verify their nonclassical features.

The measurement induced approach towards implementing
the demanding cubic nonlinearity on any input state
relies on a highly squeezed version of the cubic state.
From this resource state,
the nonlinearity is pasted onto the target input state
using the continuous-variable measurement
and quantum feed-forward control \cite{cubic,Ghose07,Sasaki06}.
A preparation of such the high-cubic state
is currently an impossible ordeal,
as it requires the cubic nonlinearity
which is inaccessible in the first place.
A recent proposal therefore suggested
use of an \emph{approximative} weak cubic state,
described as a superposition of Fock states
$\ket{0}$,  $\ket{1}$ and  $\ket{3}$ \cite{Marek11}.
The first step towards understanding and realization of
the cubic nonlinearity therefore lies in obtaining
firm grasp of the nonclassical properties
of the weak cubic state.
In this Letter, we present the first experimental preparation
of a non-Gaussian quantum state of light,
with key features consistent with state produced by cubic nonlinearity.

%\emph{Cubic state} -
The ideal cubic state,
which can be used as a resource for the nonlinear cubic gate,
can be expressed as $\int e^{i \chi_0 x^3} \ket{x}\,dx$.
Note that normalization factors are omitted in this letter
unless otherwise noted.
The cubic state can be obtained by applying cubic nonlinear interaction
$\hat{U}(\chi_0) = \exp(i \chi_0 \hat{x}^3)$ to an infinitely squeezed state.
This is a nonphysical state possessing strong nonlinear behavior,
much akin to a state with infinitely large squeezing.
Due to general inaccessibility of a cubic nonlinear operation,
any physical realization of the state needs to be some kind of approximation.
For weak cubic nonlinearity and finite energy,
the state can be approximated by
$\hat{S}(-r) (1 + i \chi \hat{x}^3) \ket{0}$ \cite{Marek11}.
Here, the cubic nonlinearity $\chi$ is given by $\chi = \chi_0 e^{3 r}$,
and $\hat{S}(-r) = \exp[-(i r/2)(\hat{x} \hat{p} + \hat{p} \hat{x})]$
is a squeezing operation - a Gaussian operation,
which can be considered feasible and well accessible
in the contemporary experimental practice \cite{mio1,mio2,mio3,mio4}.
The squeezing operation does not affect the cubic behavior of the state
and can be therefore omitted in our first attempts
to implement the cubic operation.
The approximative weak cubic state
can be then expressed in the Fock space as
\begin{equation}
\label{Eq-approxCPS}
(1 + i \chi \hat{x}^3) \ket{0} =
\ket{0} + i \frac{\chi \sqrt{15}}{2 \sqrt{2}} \ket{1 \& 3},
\end{equation}
where
$\ket{1 \& 3} = (\sqrt{3} \ket{1} + \sqrt{2} \ket{3})/\sqrt{5}$.
It is a specific superposition of zero, one and three photons,
but it can be also viewed as a superposition of
vacuum $\ket{0}$ with a state $\ket{1 \& 3}$,
which in itself is an approximation of odd superposition of coherent states.
The vacuum contribution results
from the first term of the unitary evolution
$\hat{U} (\chi) \approx 1 + i \chi \hat{x}^3$.
It is an important term for the function
of the deterministic cubic phase gate,
but at the same time it masks the nonclassical features
of the state $\ket{1 \& 3}$.
%%%%%%%%%%%%%%%%%%%%%%%%%%%%%%%%%%%%%%%%%%%%%%%%%%%%%%%%%%%%%%%%%%%%%%%%%%%%%%%%
\begin{figure}[tb]
\begin{center}
\includegraphics[width=\linewidth]{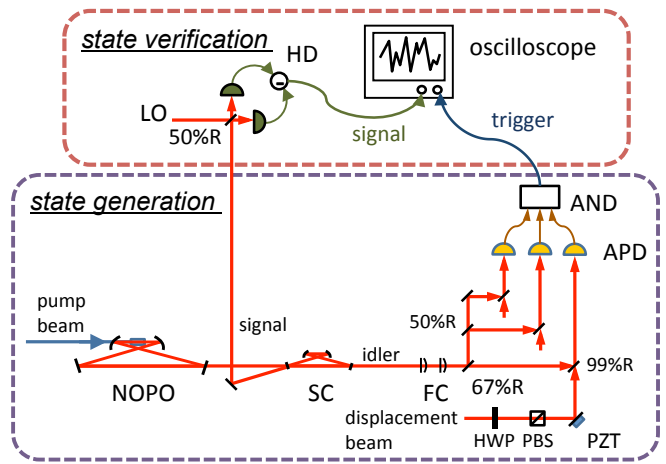}
\end{center}
\caption{%
(color online).
Experimental setup.
NOPO, non-degenerate optical parametric oscillator;
SC, split cavity;
FC, filter cavity;
HD, homodyne detector;
APD, avalanche photo diode;
HWP, half-wave plate;
PBS, polarization beamsplitter;
PZT, piezo electric transducer.
}
\label{Setup.eps}
\end{figure}
%%%%%%%%%%%%%%%%%%%%%%%%%%%%%%%%%%%%%%%%%%%%%%%%%%%%%%%%%%%%%%%%%%%%%%%%%%%%%%%%

\begin{figure}[tb]
\begin{center}
\includegraphics[width=\linewidth]{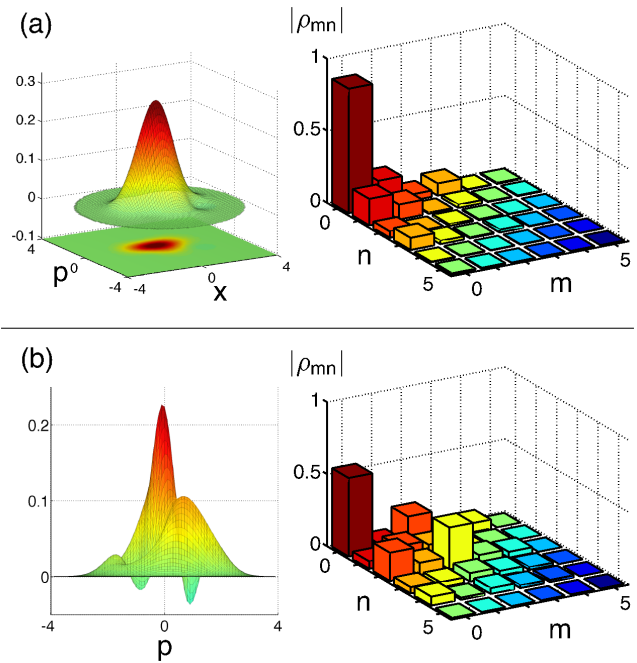}
\end{center}
\caption{%
(color online).
(a) Wigner function and density matrix of the experimentally generated state.
(b) Wigner function and density matrix of the experimentally generated state
    after a single photon is numerically subtracted from the data.
}
\label{ExpResult.eps}
\end{figure}

%\emph{Experiment} -
We attempt to generate the approximative weak cubic state of
Eq.\ \eqref{Eq-approxCPS}.
An experimental scheme is the same as the one
in the experiment of \cite{optcat3}
where superpositions of Fock states from zero to three are generated.
A schematic of the experiment is shown in Fig.\ \ref{Setup.eps}.
The light source is a continuous wave Ti:Sapphire laser of 860 nm.
With around 20 mW of pump beam of 430 nm,
a two-mode squeezed vacuum is generated
from a non-degenerate optical parametric oscillator (NOPO),
which contains a periodically-poled $\mathrm{KTiOPO_{4}}$ crystal
as an optical nonlinear media.
The pump beam is generated by second harmonic generation
of the fundamental beam,
and frequency-shifted with an acousto-optic modulator
by around 600 MHz (equal to free spectral range of NOPO, $\Delta \omega$).
As a result, photon pairs of frequency $\omega$ (signal)
and $\omega + \Delta \omega$ (idler) are obtained
($\omega$ corresponds to the frequency of the fundamental beam).
The output photons are spatially separated by a split cavity
whose free spectral range is $2 \Delta \omega$.
The idler photons passing through the split cavity
are sent to two frequency filtering cavities,
and are split into three beams with beamsplitters.
Each beam is interfered with displacement beams
at mirrors of 99\% reflectivity.
Phase of the displacement is controlled by piezo electric transducers,
and amplitude of the displacement is controlled by
rotating half-wave plates followed by polarization beamsplitters.
The idler photons are detected by avalanche photo diodes (APDs).
When APDs detect photons,
they output electronic pulses which are combined into an AND circuit
to get three-fold coincidence clicks.
The signal beam is measured by homodyne detection
with a local oscillator beam of 10 mW.
The homodyne current is sent to an oscilloscope
and stored every time of coincidence clicks.
The density matrix and Wigner function of the output state
are numerically reconstructed from a set of measured quadratures
and phases of the local oscillator beam.
The experimentally reconstructed density matrix and Wigner function
are shown in Fig.\ \ref{ExpResult.eps} (a).

%\emph{Analysis of the experimental state} -
In the following we attempt to analyze
nonlinearity inherent in the experimentally generated state.
This is not straightforward because of similarity
between the target and the vacuum states.
This makes fidelity, which is usually used in state generation scenarios,
unsuitable.
We will therefore focus on confirming
the presence of higher photon numbers in nontrivial superpositions.

We can start by applying a virtual single photon subtraction
$\hat{\rho}_\mathrm{exp} \rightarrow \hat{\rho}_{1\mathrm{sub}} =
\hat{a} \hat{\rho}_\mathrm{exp} \hat{a}^\dag
/\Tr[\hat{a} \hat{\rho}_\mathrm{exp} \hat{a}^\dag]$,
where $\rho_\mathrm{exp}$ represents the experimentally generated cubic state.
For the ideal resource state, $(1+ i\chi \hat{x}^3) \ket{0}$,
this should result in a superposition $\ket{0} + \sqrt{2} \ket{2}$,
which is a state fairly similar to
an even superposition of coherent states and, as such,
it should possess several regions of negativity.
Thus we can convert the cubic state into a state
with well known properties, which can be easily tested.
We can also remove the dominant influence of the vacuum state.
Figure \ref{ExpResult.eps} (b) shows that
the Wigner function and the density matrix of
the numerically photon-subtracted experimental state.
Notice that two distinctive regions of negativity are indeed present.
Moreover, apart from considerations involving specific states,
the areas of negativity sufficiently indicate
nonclassical behavior of the initial state,
as they would not appear
if the state was only a mixture of coherent states
($\hat{a} \ket{\alpha} \bra{\alpha} \hat{a}^\dagger \propto
\ket{\alpha} \bra{\alpha}$,
where $\ket{\alpha}$ is a coherent state with amplitude $\alpha$).
The probability of two photons $p'_2 = 0.29$ is clearly dominating
over $p'_1 = 0.12$ and $p'_3 = 0.03$,
where $p'_i = \braket{i|\hat{\rho}_{1\mathrm{sub}}|i}$.
To characterize the superposition of basis states $\ket{0}$ and $\ket{\Phi}$,
we use the normalized off-diagonal element
$\mathcal{R}_{0,\Phi}(\hat{\rho}) =
\frac{|\braket{0|\hat{\rho}|\Phi}|^2}
{\braket{0|\hat{\rho}|0} \braket{\Phi|\hat{\rho}|\Phi}}$,
which characterizes quality of any unbalanced superposition.
Since the subtraction preserves superposition of Fock states,
$\mathcal{R}_{0,2}(\hat{\rho}_{1\mathrm{sub}}) = 0.24$ after the subtraction
proves the coherent superposition
originating from the state $\ket{1 \& 3}$.
In a similar way we can confirm that the three-photon element is
significantly dominant over the two- and four-photon elements.
Two virtual photon subtractions transform the state
$\hat{\rho}_\mathrm{exp} \rightarrow \hat{\rho}_{2\mathrm{sub}} =
\hat{a}^2 \hat{\rho}_\mathrm{exp} \hat{a}^{\dag2}
/\Tr[\hat{a}^2 \hat{\rho}_\mathrm{exp} \hat{a}^{\dag2}]$,
where the single photon state is present with a probability of
$p''_1 = \braket{1|\hat{\rho}_{2\mathrm{sub}}|1} = 0.68$.
In a generated single photon state this would be
a sufficient confirmation the state cannot be emulated by
a mixture of Gaussian states.
In our case it is the argument for the strong presence of the
three-photon element.

Our analysis confirms presence
of the highly nonclassical superposition state $\ket{1 \& 3}$,
but we also need to demonstrate that the state appears in a superposition
with the vacuum state, not just as a part of mixture.
For this we look at the normalized off-diagonal element
$\mathcal{R}_{0,1 \& 3}(\hat{\rho}_\mathrm{exp})$
between the $\ket{0}$ and $\ket{1 \& 3}$
for the original (not photon-subtracted) experimental state,
which would attain the value of one for the ideal pure state,
and value of zero for a complete mixture.
In our case the value is
$\mathcal{R}_{0,1\& 3}(\hat{\rho}_\mathrm{exp}) = 0.50$,
so the superposition is present,
even if it is not perfectly visible due to effects of noise.
More importantly, the element is significantly larger than
$\mathcal{R}_{0,1\& 3^{\bot}}(\hat{\rho}_\mathrm{exp}) = 0.11$,
where $\ket{1 \& 3^{\bot}} = (\sqrt{2} \ket{1} -\sqrt{3} \ket{3})/\sqrt{5}$
is orthogonal to $\ket{1 \& 3}$.
This shows that the desired and theoretically expected superpositions
are dominant.

%%%%%%%%%%%%%%%%%%%%%%%%%%%%%%%%%%%%%%%%%%%%%%%%%%%%%%%%%%%%%%%%%%%%%%%%%%%%%%%%
%%%%%%%%%%%%%%%%%%%%%%%%%%%%%%%%%%%%%%%%%%%%%%%%%%%%%%%%%%%%%%%%%%%%%%%%%%%%%%%%

%\emph{Detecting cubic nonlinearity} -
What remains to be seen is whether the state also possesses
sufficient cubic nonlinearity which can be used for
cubic unitary transformation.
The effect of such transformation may be already visible at a classical level.
Cubic nonlinearity directly transforms
the first moments of the input state's quadratures
$\hat{x}_\mathrm{in}$ and $\hat{p}_\mathrm{in}$ according to
$\moment{\hat{x}_\mathrm{out}} = \moment{\hat{x}_\mathrm{in}}$,
$\moment{\hat{p}_\mathrm{out}} =
\moment{\hat{p}_\mathrm{in}} + 3 \chi \moment{\hat{x}_\mathrm{in}^2}$.
The first moment of $\hat{x}$ should be preserved,
while the first moment of $\hat{p}$ should become
linearly dependent on the second moment
$\moment{\hat{x}^2} = \operatorname{var}(x) + \moment{\hat{x}}^2$.
Note that $\operatorname{var}(x)$ is a variance of $\hat{x}$.
If we choose a set of input states with identical variances,
there should be observable quadratic dependence
of the first moment of $\hat{p}$ on the first moment of $\hat{x}$.

\begin{figure}
\begin{center}
\includegraphics[width=0.8\linewidth]{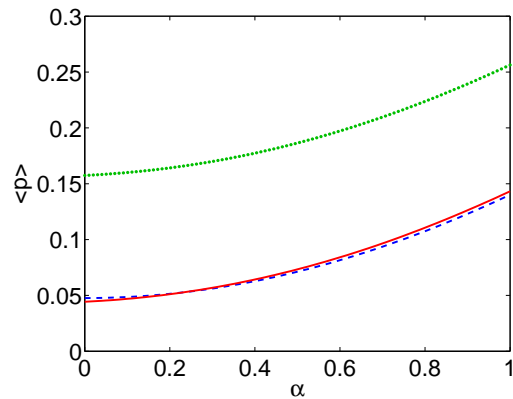}
\end{center}
\caption{%
(color online).
First moment of $p$ for various coherent states:
ideal state with $\chi = 0.090$ (dashed blue line),
experimentally generated state (dotted green line),
and experimentally generated state after the suitable displacement
$\Delta p = -0.16$ (solid red line).
}
\label{m1_comparison}
\end{figure}

We can try implementing this nonlinearity
by taking an imprint of the generated cubic state,
in a very similar manner to how a single photon can be used
to obtain a probabilistic map \cite{imprinting}.
As the set of target states
we will consider coherent states $\ket{\alpha}$,
where $0 \le \alpha \le 1$, with first moments
$\moment{\hat{x}_\mathrm{in}} = \sqrt{2} \alpha$ and
$\moment{\hat{p}_\mathrm{in}} = 0$.
The operation, imprinting nonlinearity from the ancillary mixed state
$\hat{\rho}_\mathrm{A}$ to the target state
$\hat{\rho}_\mathrm{in} = \ket{\alpha}\bra{\alpha}$ can be realized by map
\begin{equation}
\label{map}
\hat{\rho}_\mathrm{out} =
\Tr_\mathrm{A}[\hat{U}_\mathrm{BS} \hat{\rho}_\mathrm{in}
\otimes \hat{\rho}_\mathrm{A} \hat{U}^\dag_\mathrm{BS}
\ket{x = 0}_\mathrm{A} \bra{x = 0}],
\end{equation}
where $\hat{U}_\mathrm{BS}$ is a unitary operator
realizing transformation by a balanced beam splitter and
$\ket{x = 0}_\mathrm{A}$ is the zero value position eigenstate.
It can be easily confirmed that this map fuses
two states with wave functions
$\psi_\mathrm{S}(x_\mathrm{S})$ and $\psi_\mathrm{A}(x_\mathrm{A})$
into a state with wave function
$\psi_\mathrm{S}(x_\mathrm{S}/\sqrt{2})
\psi_\mathrm{A}(x_\mathrm{S}/\sqrt{2})$.
The factor $\sqrt{2}$ only introduces linear scaling of the measured data
and has no influence on any nonlinear properties.
Since the imprinting operation uses only Gaussian tools,
any non-Gaussian nonlinearity of the transformed state needs to originate in
nonlinear properties of the ancillary state $\hat{\rho}_\mathrm{A}$.
The nonlinear behavior should manifest in the first moment of quadrature $P$,
which we plot in Fig.~\ref{m1_comparison}.
We can see that the dependence is distinctively quadratic.
This behavior is actually in a very good match with
that of the ideal cubic state \eqref{Eq-approxCPS} with $\chi = 0.090$.
They only differ by a constant displacement,
which has probably arisen due to experimental imperfections
and which can be easily compensated.
This showcases our ability to prepare a quantum state
capable of imposing high-order nonlinearity in a different quantum state.

\begin{figure}
\begin{center}
\includegraphics[width=0.8\linewidth]{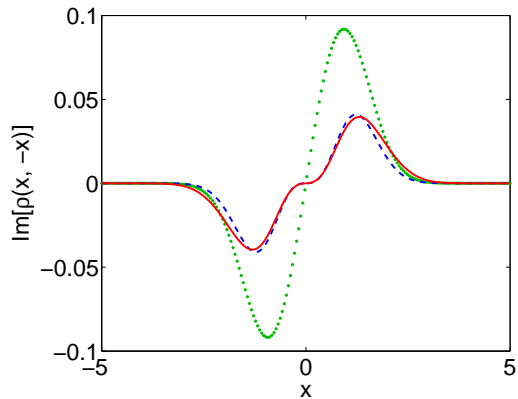}
\end{center}
\caption{
(color online).
Imaginary parts of the anti-diagonal values of coordinate density matrices
for the ideal state with $\chi = 0.090$ (dashed blue line),
the experimentally generated state (dotted green line)
and the experimentally generated state after the suitable displacement
$\Delta p = -0.17$ (solid red line).
}
\label{fig_rhoxx}
\end{figure}

We can also attempt to observe the cubic nonlinearity directly,
using density matrix in coordinate representation.
In this picture, the continuous density matrix elements are defined as
$\rho(x, x') = \braket{x|\hat{\rho}|x'}$.
The cubic nonlinearity is best visible in the imaginary part
of the main anti-diagonal: for the ideal state
$(1 + i \chi \hat{x}^3) \ket{0} \bra{0} (1 - i \chi \hat{x}^3)$,
the density matrix elements are
$\operatorname{Im}[\rho(x, -x)] =  2 \chi x^3 e^{-x^2}$
and the cubic nonlinearity is nicely visible.
One problem in this picture is that the cubic nonlinearity
can be concealed by other operations.
The second order nonlinearity does not manifest
in the imaginary part (no even order nonlinearities do),
but a simple displacement can conceal the desired behavior.
On the other hand, displacement can be quite straightforwardly
compensated by performing a virtual operation on the data.
The comparison of the ideal state, the generated state,
and the displaced generated state can be seen in Fig.\ \ref{fig_rhoxx}.
We can see that although the cubic nonlinearity is not
immediately apparent in the generated state,
the suitable displacement can reveal it effectively.
This nicely witnesses our ability to conditionally prepare
quantum state equivalent to the outcome of
the required higher-order nonlinearity.

%%%%%%%%%%%%%%%%%%%%%%%%%%%%%%%%%%%%%%%%%%%%%%%%%%%%%%%%%%%%%%%%%%%%%%%%%%%%%%%%
%%%%%%%%%%%%%%%%%%%%%%%%%%%%%%%%%%%%%%%%%%%%%%%%%%%%%%%%%%%%%%%%%%%%%%%%%%%%%%%%

%%%%%%%%%%%%%%%%%%%%%%%%%%%%%%%%%%%%%%%%%%%%%%%%%%%%%%%%%%%%%%%%%%%%%%%%%%%%%%%%
%%%%%%%%%%%%%%%%%%%%%%%%%%%%%%%%%%%%%%%%%%%%%%%%%%%%%%%%%%%%%%%%%%%%%%%%%%%%%%%%

%\emph{Summary and outlook} -
We have generated a nonclassical non-Gaussian quantum state of light,
which exhibits key features of a state
produced by unitary dynamics driven by cubic quantum nonlinearity.
Our experimental test has demonstrated the feasibility of
conditional optical preparation of the ancillary resource state
for the cubic measurement-induced nonlinearity.
As this is the first state of this kind ever to be experimentally observed,
our analysis has contributed to general understanding of
quantum states produced by the higher-order quantum nonlinearities.
This understanding is a crucial step towards physically implementing
these nonlinearities as a part of quantum information processing
and we expect first attempts in this direction to appear soon.

\emph{Acknowledgements} -
This work was partly supported by PDIS, GIA, G-COE, APSA,
and FIRST commissioned by the MEXT of Japan, and ASCR-JSPS.
K. M. acknowledges financial support from ALPS.
P. M. and R. F. acknowledge the support of the Czech Ministry of
Education under the grant LH13248.

\end{document}